\documentclass[prl,twocolumn,superscriptaddress]{revtex4-1}
\usepackage[dvipdfmx]{graphicx}
\usepackage{graphicx}
\usepackage{physics}
\usepackage{bm, amsmath, amssymb, braket}
\usepackage{times}
\usepackage{multirow}
\usepackage{ascmac}
\usepackage{amsthm}
\usepackage{float}
\usepackage{color}
\usepackage{comment}
\usepackage{mathtools}  % for math
\usepackage{mathrsfs} 
\usepackage[colorlinks=True,urlcolor=blue,linkcolor=blue,citecolor=blue]{hyperref}% add hypertext capabilities
\usepackage{xcolor}
\usepackage{multirow}
\usepackage{setspace}
\usepackage{subfigure}
\usepackage[toc,page]{appendix}

\begin{document}
\title{Artificial Gauge Field Engineered Excited-State Topology: Control of Dynamical Evolution of Localized Spinons}
\author{Jie Ren}
\affiliation{National Laboratory of Solid State Microstructures and Department of Physics, Nanjing University, Nanjing 210093, China}
\affiliation{Department of Physics, Changshu Institute of Technology, Changshu 215500, China} 
\affiliation{School of Physical Science and Technology, Suzhou University of Science and Technology, Suzhou, 215009, China}

\author{Yi-Ran Xue}
\affiliation{National Laboratory of Solid State Microstructures and Department of Physics, Nanjing University, Nanjing 210093, China}

\author{Run-Jia Luo}
\affiliation{National Laboratory of Solid State Microstructures and Department of Physics, Nanjing University, Nanjing 210093, China}

\author{Rui Wang}
\email{rwang89@nju.edu.cn}
\affiliation{National Laboratory of Solid State Microstructures and Department of Physics, Nanjing University, Nanjing 210093, China}
\affiliation{Collaborative Innovation Center of Advanced Microstructures, Nanjing University, Nanjing 210093, China}
\affiliation{Jiangsu Physical Science Research Center}
\affiliation{Hefei National Laboratory, Hefei 230088, People's Republic of China }

\author{Baigeng Wang}
\email{bgwang@nju.edu.cn}
\affiliation{National Laboratory of Solid State Microstructures and Department of Physics, Nanjing University, Nanjing 210093, China}
\affiliation{Collaborative Innovation Center of Advanced Microstructures, Nanjing University, Nanjing 210093, China}
\affiliation{Jiangsu Physical Science Research Center}

\begin{abstract}
Spinons are elementary excitations at the core of frustrated quantum magnets. Although it is well-established that a pair of spinons can emerge from a magnon via deconfinement, controlled manipulation of individual spinons and direct observation of  their deconfinement remain elusive. We propose an artificial gauge field scenario that enables the engineering of specific excited states in quantum spin models. This generates spatially localized individual spinons  with high controllability. By applying time-dependent gauge fields, we realize adiabatic braiding of these spinons, as well as their dynamical evolution in a controllable manner. These results not only provide the first direct visualization of individual spinons localized in the bulk, but also point to new possibilities to simulate their confinement process. Finally, we demonstrate the feasibility of our scenario in Rydberg atoms, which suggests an experimentally viable direction--gauge field engineering of correlated phenomena in excited states.
\end{abstract}

\maketitle
\emph{\color{blue}{Introduction.--}} Spinons are exotic quasi-particles playing a crucial role in condensed matter physics. These fractional excitations carry zero charge and half-integer spin, and are essential for understanding strongly correlated phases, such as quantum spin liquids (QSLs) \cite{Anderson1,xgwen,Broholm} and high-temperature superconductors \cite{Anderson,Bednorz,Nagaosa}. In QSLs, the spinons usually emerge through the deconfinement from magnons \cite{Hanus,Wortis}, highlighting the complex collective behavior of these systems. Recent experiments have reported indirect evidence supporting the presence of spinons. For example, transport measurements have shown spin-charge separation in Luttinger liquids \cite{Jompol} and neutron/Raman scattering has revealed broad spectral features in spin liquid candidates \cite{thhan,palee1,Knolle}. Despite these progresses, direct observation of \textit{individual} spinons has remained elusive for decades.

\begin{figure}[t]
    \centering
    \includegraphics[width=\linewidth]{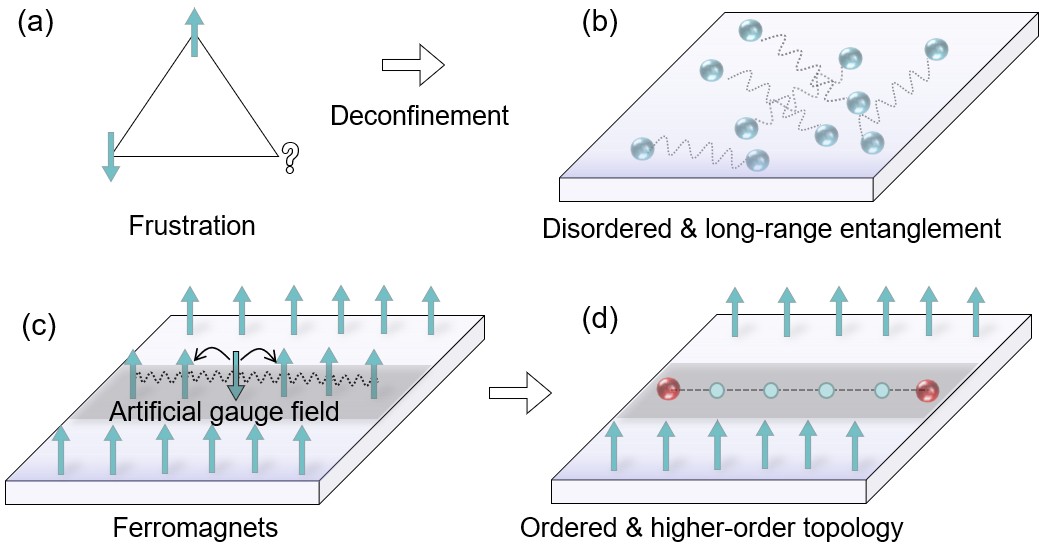}
    \caption{ (a) and (b) illustrate the deconfinement of spinons in quantum spin liquids driven by magnetic frustration. (c) (d) schematically indicate the gauge field scenario to engineer excited states in quantum spin models. The gauge field induces higher order topology and localized spinons (marked by red) with high controllability. }
    \label{fig1}
\end{figure}

Observation and manipulation of individual spinons, particularly in two dimensions, are of importance because they would offer new insights into topological orders \cite{xgwen1,xgwen2,xiechen}, anyon statistics \cite{Wilczek,Bartolomei,Halperin}, advancing potential applications in quantum computation \cite{Nayak,ayukitaev,Semeghini}. However, capturing a single spinon out of a QSL material is extremely challenging \cite{wzhu,yfu,mpotts}, because of their itinerant nature in disordered, long-range entangled ground state (Fig.\ref{fig1}(a)(b)). This raises a critical question, i.e, is it possible to spatially “localize” a single spinon in two-dimensions in a controllable manner, which enables further engineering of its dynamics, including the deconfinement-confinement and the braiding statistics?

In this Letter, we solve this key problem by proposing an artificial gauge field (AGF) engineering mechanism. AGF has recently been found to efficiently modulate the ground state properties of topological photonic \cite{Cheng, Umucalilar,Kejie} and phononic crystalline \cite{Mathew,xwen,Pengtao,liluo,qiyun,Zhaoju} systems. Here, we show that it can even induce high-order topologies \cite{Schindler,Schindler2,Christopher,haoran} in the \textit{excited states} of quantum spin systems, enabling stabilization of localized spinons and control of their dynamical evolutions.  We demonstrate this mechanism by applying AGF to a line of local bonds of a quantum spin model that supports topological magnons, as shown by Fig.\ref{fig1}(c). Our spin wave theory predicts that the AGF imprints an effective 1D topological magnon insulator embedded within the 2D bulk, generating higher-order topological modes at the line boundaries (Fig.\ref{fig1}(d)).  

Using  density matrix renormalization group (DMRG)-based approaches, we unambiguously observe that the low-energy excited state, i.e., a magnon, is fractionalized into a pair of boundary modes. Remarkably, the boundary modes exhibit spin-1/2 and fermionic statistics, validating that they are essentially spinons. Unlike the frustration-induced itinerant spinons in QSLs (Fig.\ref{fig1}(a)(b)), the spinons here emerge as second-order boundary modes \cite{Christopher,haoran,congchen}, thus are spatially localized (Fig.\ref{fig1}(d)). The locality enables additional dynamic control. By manipulating time-dependent AGFs, we realize novel dynamical evolutions, including their adiabatic braiding and delocalization into the bulk. Interestingly, the latter points to a novel way to simulate their confinement where a pair of spinons merge into a magnon mode. Furthermore, we discuss experimental feasibility in programmable platforms such as Rydberg atom arrays \cite{Maskara,Bluvstein}. Our work presents the first numerical visualization of isolated spinons in quantum spin systems and opens new possibilities to simulate their confinement, establishing the AGF engineering as a promising paradigm for manipulating exotic excited-states.

\textit{\color{blue}{AGF modulated excited state topology.--}} We consider a quantum spin model decorated by a classical $\mathrm{Z}_2$ field  on a honeycomb lattice, i.e.,
\begin{equation}\label{eqm1}
\begin{split}
H&=\sum_{\langle i,j\rangle}J_{ij}(t)(S^x_{i}S^x_{j}+S^y_{i} S^y_{j})+\sum_iB_iS^z_i\\
&-K\sum_{\langle i,j\rangle}S_i^zS_j^z+D\sum_{\langle\langle i,j\rangle\rangle}\nu_{ij}\mathbf{\hat{z}}\cdot(\mathbf{S}_{i}\times\mathbf{S}_{j}),
\end{split}
	\end{equation}
where the Heisenberg coupling $J_{ij}(t)=-J\sigma^z_{ij}(t)$, and $\sigma^z_{ij}(t)$  is the $Z_2$ field defined on the nearest-neighbor bond. $\sigma^z_{ij}$ takes values $\pm1$ and can be time-dependent.  Note that the first term in Eq.\eqref{eqm1} is formally invariant under the combined operation, i.e., $S^{x,y}_i\rightarrow-S^{x,y}_i, \forall i$, and the local $Z_2$ gauge transformation, in consistence with lattice gauge theories coupled to matter fields. Thus, the classical field here mimics a $Z_2$ gauge field without intrinsic dynamics and is therefore called an AGF.  The last term is the  Dzyaloshinskii–Moriya (DM) interaction with $\nu_{ij}=\pm1$, determined by the clockwise or counterclockwise direction of the vectors connecting the next-nearest sites. The local magnetic field $B_i$ offers another tuning knob. 

For the gauge field free version of Eq.\eqref{eqm1}, i.e., $J_{ij}(t)=-J$,  the linear spin-wave theory suggested the emergence of topological magnons \cite{Kim}. However, whether the topology is robust against magnon-magnon interactions is still under debate \cite{Habel}. Here, we consider a large Ising interaction $K$ that favors the ferromagnetic ground state along $z$-direction, which suppresses the magnon-magnon interactions (Sec.5 of the supplemental materials). In this way,  we will provide  smoking gun numerical evidence  that supports topological magnons beyond the spin-wave approximation.  Moreover,  our focus here is on the effects of the AGF, which will be shown to induce new topological phenomena in the \textit{excited states}. Eq.\eqref{eqm1} points to a largely unexplored area, i.e., the combination between AGF and excited-state physics.

\begin{figure}[t]
    \centering
    \includegraphics[width=\linewidth]{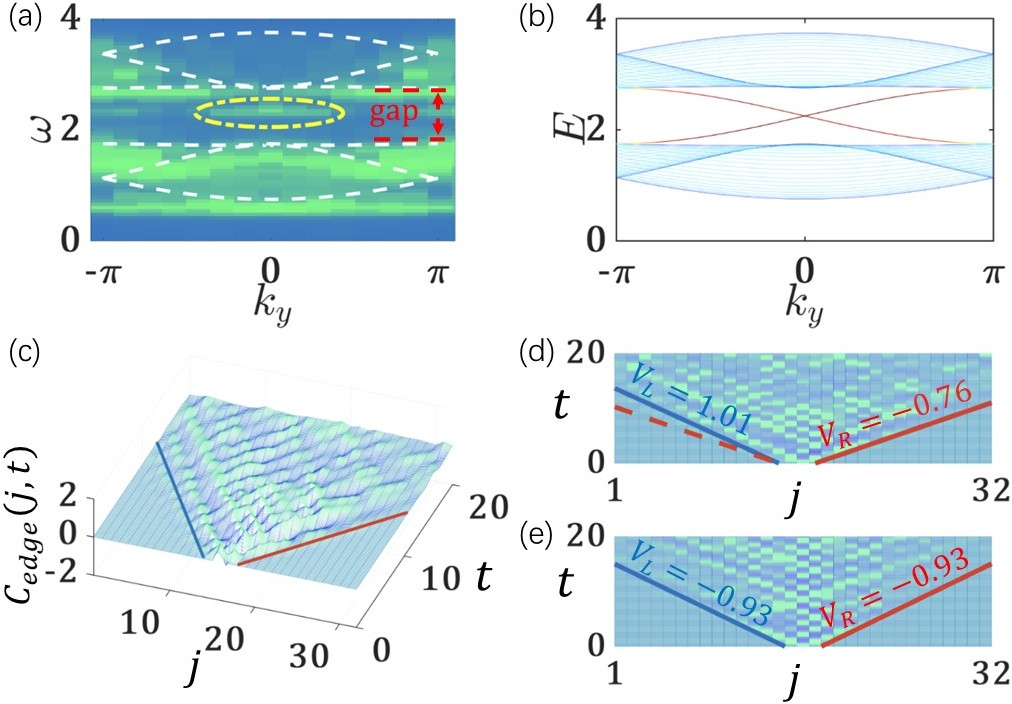}
    \caption{(a) The color plot of $S(q_y,\omega)$ obtained by TDVP on a cylinder geometry where the PBC and OBC are adopted along $y$ and $x$-direction, respectively.  The system size, $N_x=24$, $N_y=16$, is used. The dashed circle highlights signatures of in-gap states. (b) The magnon dispersion obtained by spin-wave theory, which is also plot by the dashed white curves in (a) for comparison. (c) shows the edge correlation function $C_{\mathrm{edge}}(j,t)$. The left and right propagation of the spin excitations are highlighted by blue and red lines, respectively. (d) indicates the different velocities between the left and right movers, implying chiral nature of the edge state. This is however absent for quantum spin models supporting topologically trivial magnons (e). $K=1.5$, $D=0.2$, $J=1$ are used for (a)-(d). }
    \label{fig2}
\end{figure}

We first solve the gauge field free model nonperturbatively, using the large-scale time-dependent DMRG method \cite{sup}. By using time-dependent variational principle (TDVP) in the manifold of matrix product states \cite{Ulrich02,Haegeman,Fishman,sup}, we calculate the spin correlations $C(j,t)=\langle S^+ _{N/2} (0)S^-_{N/2+ j} (t)\rangle$ on a finite size lattice with periodic boundary condition (PBC) along $y$-direction.  $S^{\pm}$ is the spin-raising and lowering operator, $N$ is the  number of sites, and $N/2$ denotes the central site. From the Fourier transformation of $C(j,t)$, i.e., the dynamical structure factor $S(q,\omega)$, we can read off the  magnon dispersion. As shown by Fig.\ref{fig2}(a), a significant magnon gap is opened around $1.7\lesssim\omega\lesssim2.6$,  as a result of the DM interaction. The gap and the dispersion quantitatively agree with those obtained by the linear spin wave theory (Fig.\ref{fig2}(b)). The latter predicts a topological magnon insulator with chiral edge states characterized by Chern number $C=1$.

Notably, signatures of in-gap excitations are also found in Fig.\ref{fig2}(a), as marked by the yellow dashed circle. The spin wave results suggest that these may originate from the edge states.  However, because the edge contribution to $S(q,\omega)$ is weak compared to those of the bulk, the signature here is not clear enough. Thus, we further calculate the edge spin correlation function, $C_{\mathrm{edge}}(j,t)=\langle S^+ _{i} (0)S^-_{i+j} (t)\rangle$, where  $i$ and $i+j$ denotes the sites on the boundary. $C_{\mathrm{edge}}(j,t)$ is essentially the propagator of the spin excitation along the edge, which should reflect the key features of edge states if they do exist. 

From  $C_{\mathrm{edge}}(j,t)$ shown in Fig.\ref{fig2}(c), we clearly observe the left and right propagation of the spin excitations along the edge, highlighted by the blue and red lines. Furthermore, it is found that the left and right movers  exhibit different velocities, $v_{L}\neq v_R$ ( Fig.\ref{fig2}(d)).  For comparison, we also plot $C_{\mathrm{edge}}(j,t)$ for a topologically trivial insulator without edge states. Replacing the DM interaction by a sublattice-staggered magnetic fields leads to a trivial magnon insulator with $C=0$. In this case, $v_{L}=v_R$ is observed (Fig.\ref{fig2}(e)). Therefore, $v_{L}\neq v_R$ for $D\neq0$ in  Fig. \ref{fig2}(d) reveals the chiral nature of the edge state, suggesting the robustness of  topological magnons. 

\begin{figure}[t]
    \centering
    \includegraphics[width=\linewidth]{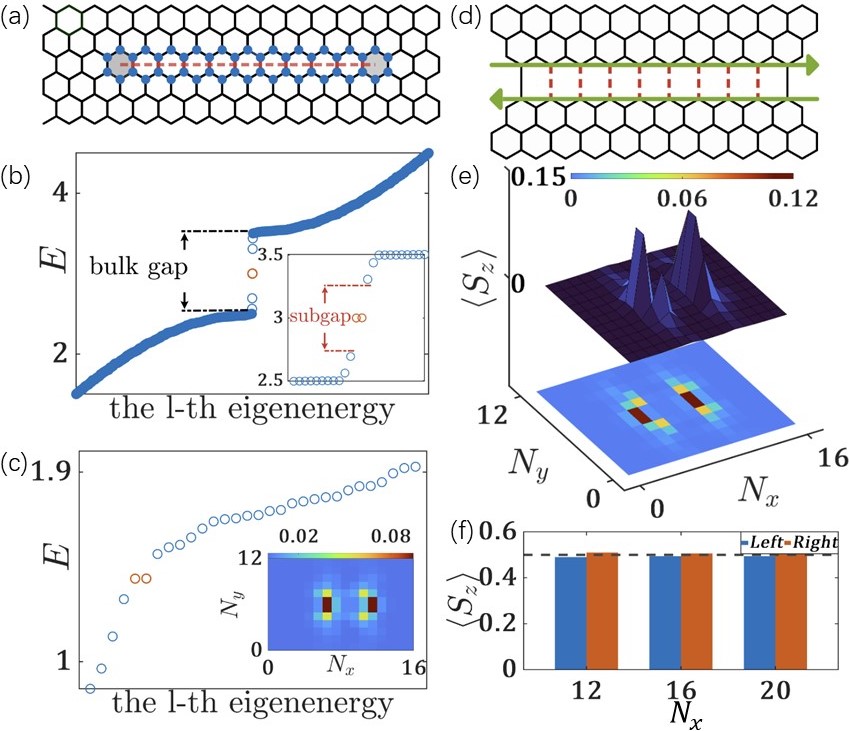}
    \caption{(a) The $Z_2$ lattice gauge field constitute a flux line defined on the quantum spin model.  (b) Spin-wave results of the eigenenergies for Eq.\eqref{eqm1} with a static $Z_2$ gauge field. The red circle denotes degenerate in-gap topological modes. (c) The same as (b) but with further turning magnetic fields, $B=-0.5$, on the sites around the flux line. The inset shows the spatial distribution of the in-gap topological modes. (d) Schematic plot of the edge-reconnection picture accounting for the occurrence of edge modes. (e) shows the wave-function distribution of $|ES_5\rangle$ obtained by DMRG. The distribution for the $|ES_6\rangle$ state is the same. For demonstration, the length of flux line is set to $L=4$ in (b),(c),(e).  (f) shows the total $\langle S^z\rangle$ around the left and right ends of the flux line for different system sizes. $K=2$, $D=0.2$, $J=1$ are used for all figures. The system size is 16$\times$12 in (b)(c)(e)}.
    \label{fig3}
\end{figure}

We now turn on a static $Z_2$ AGF, $J_{ij}(t)=-J\sigma^z_{ij}$. Different from previous works on the ground state topology \cite{lbshao,yxzhao1,haoranx,zychen,Luscher}, here we  focus on the novel excited states induced by the AGF.  As illustrated by Fig.\ref{fig3}(a),  we let $\sigma^z_{ij}=-1$ for the bonds $\langle ij\rangle$  crossing the horizontal dashed line and  $\sigma^z_{ij}=1$ for the bonds elsewhere.  The $\sigma^z_{ij}=-1$ bonds constitute a flux line embedded in the bulk.  At the spin-wave level, the eigenenergies are obtained and plot in Fig.\ref{fig3}(b). Despite the bulk gap, we  observe a number of in-gap modes originating from the edge states. Interestingly, a subgap is opened in the edge states by the gauge field, generating a pair of ``isolated" modes located at the subgap center (the red circles). 

There is a pure topological origin for the isolated modes. Following the reconnected edge picture in Ref.\cite{dhlee}, the $Z_2$ gauge field in Fig.\ref{fig3}(a) can be viewed as a line defect connecting two topological magnon insulators, as shown in Fig.\ref{fig3}(d). Thus, two edge states with opposite chiralities emerge at the boundaries, described by $H_{\mathrm{edge}}=-iv_F\int dx \psi^{\dagger}(x)\tau^z\partial_x\psi(x)$. On the flux line, the two magnonic edge states are reconnected by sign-reversed couplings. This leads to a spatially dependent mass term, $H_{\mathrm{mass}}=\int dx\psi^{\dagger}(x)m(x)\tau^x\psi(x)$, with two domain walls at the line boundaries. The mass term opens a gap in the edge, generating a pair of in-gap topological modes (Fig.\ref{fig3}(b)). The two modes form a two-fold degenerate Hilbert space, akin to that of Majorana fermions in p-wave superconductors \cite{ykitaev}. In analogy with the charge fractionalization in the polyacetylene \cite{Jackiw,Schrieffer}, the magnons here should also display fractionalization, but in terms of the spin quantum number.

The real challenge here lies in how to observe the fractionalization non-perturbatively beyond the spin-wave picture. To this end, we utilize the tuning knob $B_i$ defined in Eq.\eqref{eqm1}. By turning on magnetic fields on the sites surrounding the flux line (indicated by the blue dots in Fig.\ref{fig3}(a)),  the eigenenergies corresponding to the edge states and the topological modes can be significantly lowered, as shown by Fig.\ref{fig3}(c). This renders their numerical observation possible by examining the first several excited states. 

Based on the DMRG calculations on a finite sized lattice under the torus geometry, we obtain the lowest excited states and extract their $S^z$ expectations.  For  $K>J$, the ground state $|GS\rangle$ is a ferromagnet ground state with $\langle GS| S^z|GS\rangle\sim-1/2$ throughout the lattice. Then, the magnon excitations on top of the ground state are captured by, $\Delta S^z_i\equiv\langle ES_n| S^z_i |ES_n\rangle-\langle GS|S^z_i|GS\rangle$, where $|ES_n\rangle$ denotes the $n$-th excited states. Our calculated excited states all satisfy $\Delta S^z_{\mathrm{tot}}=\sum_i\Delta S^z_i=1$, describing spin-1 excitations.  Notably, we find that  $|ES_5\rangle$ and  $|ES_6\rangle$ are energetically degenerate, with $E_5\simeq E_6=-137.121$ \cite{footnoteadd}. Their $\Delta S^z_i$ distributions are also the same, as shown by Fig.\ref{fig3}(e).  Remarkably, they both exhibit two separate localized modes around the two ends of the flux line. The result is in well agreement with that obtained by spin-wave theory, i.e.,  inset to Fig.\ref{fig3}(c). More importantly, $\Delta S^z\sim1/2$ is numerically extracted for both the left and right mode, independent of system sizes (Fig.\ref{fig3}(f)). This clearly demonstrates the emergence of localized spin-1/2 excitations, fractionalized from a spin-1 magnon.   As discussed above, these fractionalized modes are the edge states of an 1D topological magnon insulator embedded in 2D topological magnons. Thus, we essentially find that the AGF endows a second-order topology into excited states of the quantum spin model.  The stability of the localized edge modes are further analyzed in Sec.5 of Supplemental Materials.

%Interestingly, instead of a spatially extended magnon mode (with a certain momentum), we find two localized modes around the two ends of the flux line (the dashed line of Fig.XX).  Remarkably, the total $\langle S^z\rangle$ of each localized mode is found to be exactly $\sim1/2$. Thus, we numerically observe fractionalization of the spin-1 magnon into two spin-1/2 localized modes, as a result of the modulation of $Z_2$ gauge field. For the second and third excited state...

%In order to perform the Fourier transform in time in (S(q,w) ) one has to deal with the problem that numerical data are only available for a finite time interval t ∈ [0, tmax]. The maximal simulation time t_max ∼ 40, and these data are then extended in time by using linear prediction [3]，which can lead to a smooth exponentially decaying extrapolation of the data[4,5]. 

\begin{figure}[t]
    \centering
    \includegraphics[width=\linewidth]{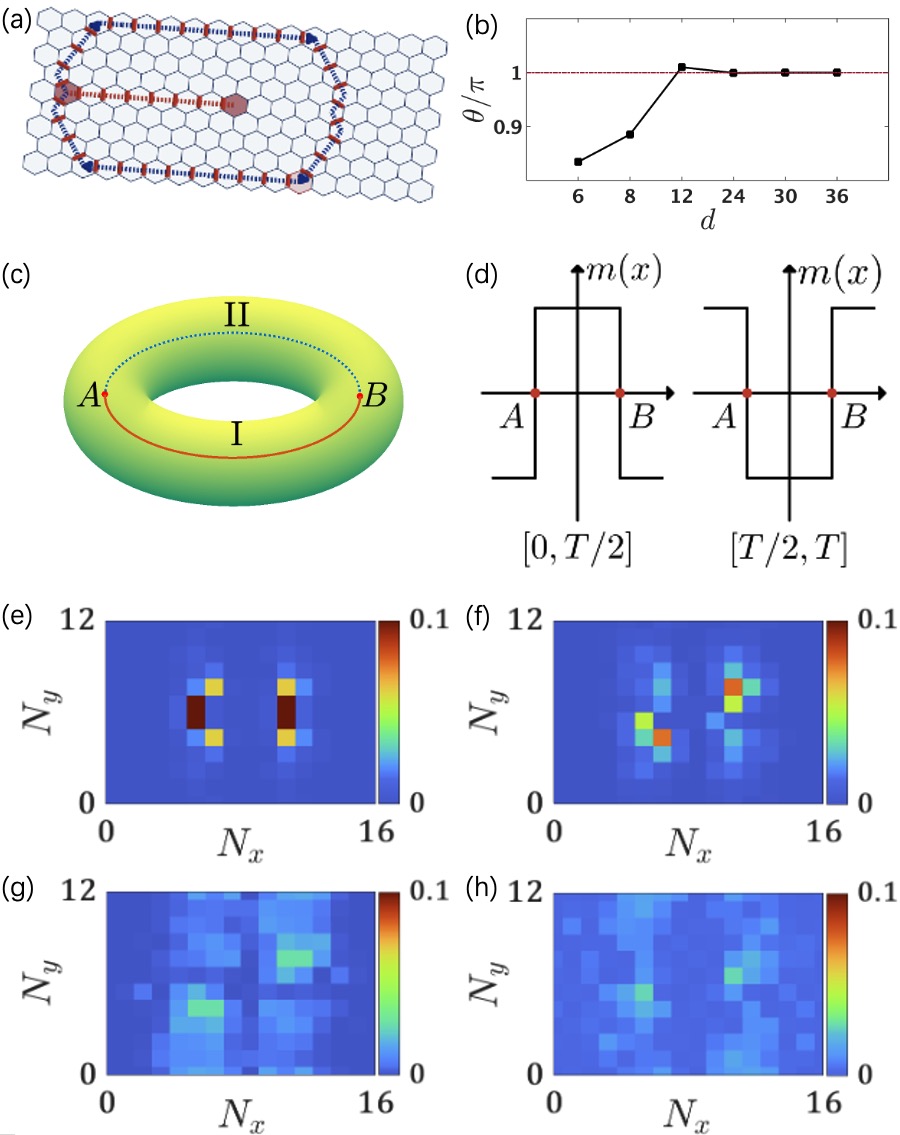}
    \caption{ (a) indicates the dynamic control of the AGF which induces the braiding of the two localized states. (b) The calculated statistical angle $\theta$ for different loops of radius $d$. (c) indicates the alternating gauge field applied on the torus, which generates two different effective mass terms sharing the same domain walls shown in (d). (e)-(h) show the evolution $\langle S^z_{i}\rangle(t)$ starting from the initial state $|ES_5\rangle$, obtained by time-dependent DMRG. $t=0,4,8,24$ for (e),(f),(g),(h), respectively. $T=0.2$ and all other parameters are the same with Fig.\ref{fig3}.}
    \label{fig4}
\end{figure}

\textit{\color{blue}{Spinon braiding and dynamical evolutions.--}} 
We now step into the  AGF-induced dynamical phenomena of the excited state. We first investigate the braiding of the fractionalized modes \cite{sup}.  These modes can be  made mobile by deforming the flux line,  as illustrated by Fig.\ref{fig4}(a).  Adiabatically moving one end (denoted by A) along a loop trajectory enclosing the other (B) gives rise to a geometric phase, $\theta_1$. Similarly, a different phase, $\theta_2$, will be accumulated for the another trajectory but not enclosing B \cite{sup}. The statistical angle $\theta$ , i.e., half of the Berry phase accumulated in the braiding, is then given by $\theta=(\theta_1-\theta_2)/2$ \cite{yingran}. We regard the braiding as a number of successive evolution steps and calculate the phase accumulation in each step. The sum of the phases for all steps leads to the statistical angle shown in Fig.\ref{fig4}(b). For small loops, the adiabaticity is not well preserved, resulting in deviations from the quantized $\theta$. For sufficiently large loops,  the statistics exactly approaches $\theta=\pi$. Hence, the topological spin-1/2 modes are fermionic, i.e., spinons.  

Then we explore more exotic dynamical evolutions of the spinon modes via engineering of AGF.  We note that, on the torus, there are two  gauge-inequivalent AGF configurations,  $|\{\sigma^z_{ij}\}_{\mathrm{I}}\rangle$ and $|\{\sigma^z_{ij}\}_{\mathrm{II}}\rangle$, both supporting  spinons at the same ends (A and B), as  indicated by I and II in Fig.\ref{fig4}(c). Both the two  configurations have a flux line with the length $N_{\mathrm{line}}$, and are energetically degenerate.  Then, we turn on  time-periodic oscillations between the two configurations.  For  $t\in[0,T]$ and along the total line (I+II), the AGF modulated interactions are given by \cite{note1}, 
\begin{equation}\label{eqm2}
J_{ij}(t)\propto
\begin{cases}
~~~2\Theta(T/2-t)-1, ~~\text{if $\langle i,j\rangle\in$ I}\\
1-2\Theta(T/2-t),  ~~\text{if $\langle i,j\rangle\in$ II}
\end{cases}
\end{equation}
where $\Theta(x)$ is the Heaviside step function. Correspondingly, this generates two  oscilating Hamiltonians, $H_{\mathrm{I}}$ and $H_{\mathrm{II}}$, in a period. The oscillation could simulate the effect of gauge fluctuations as will be clear in the following.

The dynamics of  Eq.\eqref{eqm1} under Eq.\eqref{eqm2}  can be investigated using time-dependent DMRG.  The initial state $|\Psi(0)\rangle$ is prepared as $|ES_5\rangle$, which has been shown to support two spinons at the two ends A and B. We then calculate the evolution under Eq.\eqref{eqm2} by examining, $\langle S^z_i\rangle(t)=\langle \Psi(t)|S^z_i|\Psi(t)\rangle$ \cite{sup}. As shown by Fig.\ref{fig4}(e)-(h), the two initially localized spinons become more and more delocalized. With increasing $t$, they gradually spread into the bulk, and can no longer be distinguished from each other (Fig.\ref{fig4}(h)). The final steady state at long times has little imprints of the initial state (Sec.7 of the supplemental materials), which describes a spin-1 excitation, i.e., a magnon,  shared by many sites over the lattice. 

The spreading phenomenon of the spinon modes is nontrivial. It is essentially driven by nonlocal perturbations. For $t\in[T/2,T]$, the evolution of the initial state  $|ES_5\rangle$ (an eigenstate of $H_{\mathrm{I}}$) is governed by the operator, $e^{-iH_\mathrm{II}t}=e^{-iH_\mathrm{I}t+i(H_{\mathrm{I}}-H_{\mathrm{II}})t}$. Note that the term, $H_{\mathrm{I}}-H_{\mathrm{II}}$, involves exchange interactions along the whole closed loop on the torus, and is thus nonlocal.  The evolution under such a nonlocal operator results in the spreading of localized spinons. In sharp contrast, under driving by local perturbations, the spinons remain intact and localized at A and B, as shown by Sec.5 of the Supplemental Materials. The fact that localized spinons  only become delocalized under nonlocal perturbations clearly indicates the topological nature of these created excitations. 

The spreading phenomenon of spinons can be understood using the Floquet theory  \cite{Eckardt,Bukov}. Following the reconnected edge picture above, the excitations on the total flux line (I+II) is described by an effective 1D edge theory under PBC, $H_{\mathrm{tot}}(t)=H_{\mathrm{edge}}+H_{\mathrm{mass}}(t)$, i.e.,
\begin{equation}
H_{\mathrm{tot}}(t)=\int dx \psi^{\dagger}(x)[-iv_F\tau^z\partial_x+m(x,t)\tau^x]\psi(x),
\end{equation}
where $m(x,t)$ is a time-dependent mass term supporting domain walls at A and B. In a period, $m(x,t)$ oscillates between the two configurations in Fig.\ref{fig4}(d). In the high frequency expansion, the stroboscopic evolution is determined by the Floquet Hamiltonian \cite{Anisimovas}, $\mathcal{H}_F\simeq\mathcal{H}_{\mathrm{tot},0}+[\mathcal{H}_{\mathrm{tot},1},\mathcal{H}_{\mathrm{tot},-1}]/\omega$, where $\mathcal{H}_{\mathrm{tot},n}=\frac{1}{T}\int^T_0 e^{in\omega t}[v_Fk\tau^z+m(x,t)\tau^x]$ and $\omega=2\pi/T$. The commutator here vanishes. Besides, for $\mathcal{H}_{\mathrm{tot},0}$, the integral of the mass term $m(x,t)$ also vanishes in a period due to the cancellation between I  and II.  As a result, the Floquet Hamiltonian determining the steady state is obtained as, $\mathcal{H}_F=\int dx \psi^{\dagger}(x)[-iv_F\tau^z\partial_x]\psi(x)$. Clearly, it no longer supports fractionalized modes but only the original magnons, accounting for the numerically observed spinon delocalization. 

More importantly, the Floquet engineering of AGF proposed above  points to a novel way to simulate the fluctuation of gauge field, opening new possibilities to observe spinon confinement \cite{wenbook}.  Considering a series of AGF configurations, $|\{\sigma^z_{ij}\}_{l}\rangle$ (corresponding to the Hamiltonian $H_l$) with $l=1,..N_c$ in a driving period, the dominating zeroth-order Floquet Hamiltonian  is then obtained in the high frequency regime as, $H_{\mathrm{tot},0}=\lim_{T\rightarrow0}\frac{1}{T}\int^T_0dtH(t)=\lim_{T\rightarrow0}\frac{1}{T}[\int^{T/N_c}_0dtH_{1}+...+\int^{T}_{(N_c-1)T/N_c}dtH_{N_c}]$. This indicates that the AGF configurations  $|\{\sigma^z_{ij}\}_l\rangle$  are effectively averaged in a period, resulting in the superpositioned configuration,  $|\Psi\rangle=\sum ^{N_c}_{l=1}|\{\sigma^z_{ij}\}_l\rangle$ . This mimics the gauge fluctuations, which essentially describe superposition of different gauge configurations, thereby suggesting a new approach to simulate the confinement in an non-equilibrium fashion. 

%Although it is known in the spin liquid context that a pair of spinons can be confined into a magnon by gauge fluctuations, the direct numerical observation has been lacking. This can be simulated using our scenario.  Despite the close analogy between our studied model and the effective field theory of a $Z_2$ spin liquid \cite{sup}, the AGF here  does not have an intrinsic dynamics. The dynamics can result in fluctuations between different gauge configurations, leading to confinement \cite{sup}. Thus, to simulate the confinement, a natural way is to introduce time-fluctuating $Z_2$ field $\sigma^z_{ij}(t)$, encompassing configurations that are gauge inequivalent. 

Based on the above scheme, we further consider a Floquet driving protocol whose period consists of a number of AGF configurations connected by local $S^x$ operators (Sec.9 of Supplemental Materials). Interestingly, time-dependent DMRG calculations reveal signatures of confinement, where the two localized spinon modes are brought closer to each other and merge into a localized magnon mode. Besides, assuming a dynamic term for the $Z_2$ field,  the energy cost to create a pair of spinons on top of the driven state is found to be proportional to $N_{\mathrm{line}}$, i.e., the distance between of the two modes. This is the defining feature of confinement, implying new possibilities to simulate spinon confinement via engineering of AGF.

\textit{\color{blue}{Discussion and conclusion.}}--Our scenario can be realized using Rydberg atoms \cite{sup}.  We consider trapped Rydberg atom arrays forming a honeycomb lattice. Each atom is excited to a two-state manifold that are of the same parity. The second-order perturbation mediates an XXZ type interaction between a pair of two atoms \cite{Whitlock,Kunimi}.  Considering  up to the next-nearest neighbors,  this realizes a 2D XXZ model with a magnetic field on the honeycomb lattice. 

Recently, local modulations of Rydberg atoms have been shown to generate a relative phase $\theta_{ij}$ between the $S^+_i$ and $S^-_j$ of neighboring atoms $i$ and $j$  \cite{Naveen}. Possible local modulations involve the manipulations of the interaction energy \cite{Roushan} or the detuning of the energy levels associated with a local atom \cite{dwwang}. The latter has recently been implemented via AC Stark shifts of Rydberg atoms \cite{Barredo}, where the generated local relative phase is $\theta_{ij}=\Delta\omega_0\tau$, with $\Delta\omega_0$ being the controllable light-shift on the local state and $\tau$ being the addressing time. On the honeycomb lattice, relative local phases between intra-sublatice and inter-sublattice sites can be induced, which simulates an effective Hamiltonian sharing the same form as Eq.\eqref{eqm1} \cite{sup}.  Then, the localized spinons predicted here could  be detected via the excitation spectroscopy of Rydberg atoms \cite{Peyronel}, fluorescence imaging or site-resolved measurements \cite{Endres,Okuno,Kwona}.

Our proposed AGF mechanism may be also applied to produce other topological excitations, e.g., Majorana fermions, and enhance their controllability. Our work therefore points to an efficient approach to modulate various excited-state phenomena, ranging from topological excitons \cite{rwang} to quantum scars \cite{Serbyn}.

R. J. and Y. X. contribute equally to this work.
\begin{acknowledgments}
R. W. acknowledges Tigran Sedrakyan for fruitful discussions. This work was supported by the Innovation Program for Quantum Science and Technology (Grant no. 2021ZD0302800), the National R\&D Program of China (2022YFA1403601), the National Natural Science Foundation of China (No. 12322402, No. 12274206), the Natural Science Foundation of Jiangsu Province (No. BK20233001), and the Xiaomi foundation.
\end{acknowledgments}


\begin{thebibliography}{99}

\bibitem{Anderson1} P. W. Anderson,  Mat. Res. Bull. \textbf{8}, 153-160 (1973). 

\bibitem{xgwen} X. G. Wen, Phys. Rev. B \textbf{44}, 2664 (1991).

\bibitem{Broholm} C. Broholm, R. J. Cava, S. A. Kivelson, D. G. Nocera, M. R. Norman and T. Senthil, \textit{Science}  \textbf\textbf{{367}}, eaay0668  (2020).

\bibitem{Anderson}P. W. Anderson, Science \textbf{235}, 1196(1987).

\bibitem{Bednorz} J. G. Bednorz and K. A. M\"{u}ller,  Possible high $T_c$ superconductivity in the Ba-La-Cu-O system, Z. Physik B Condensed Matter \textbf{64}, 189 (1986). 

\bibitem{Nagaosa} P. A. Lee, N. Nagaosa, and X.-G. Wen, Rev. Mod. Phys. \textbf{78}, 17 (2006).

\bibitem{Hanus} J. Hanus, Phys. Rev. Lett. \textbf{11}, 336 (1963). 

\bibitem{Wortis} M. Wortis, Phys. Rev. \textbf{132}, 85 (1963).

\bibitem{Jompol}Y. Jompol, C. J. B. Ford, J. P. Griffiths, I. Farrer, G. A. C. Jones, D. Anderson, D. A. Ritchie, T. W. Silk, A. J. Schofield, Science \textbf{325} 597-601 (2009). 

%%%%%%  
\bibitem{thhan} Tian-Heng Han, Joel S. Helton, Shaoyan Chu, Daniel G. Nocera, Jose A. Rodriguez-Rivera, Collin Broholm and Young S. Lee, Nature  \textbf{492},  406 (2012).

\bibitem{palee1} P. A. Lee, Science \textbf{321}, 1306-1307 (2008).

\bibitem{Knolle} J. Knolle, Gia-Wei Chern, D. L. Kovrizhin, R. Moessner, and N. B. Perkins, Phys. Rev. Lett. \textbf{113}, 187201 (2014). 

%%%%%% topological orders 

\bibitem{xgwen1} X. G. Wen, Adv. Phys. \textbf{44}, 405 (1995).

\bibitem{xgwen2} X.  G. Wen, Rev. Mod. Phys. \textbf{89}, 041004 (2017).

\bibitem{xiechen} X. Chen, Zheng-Cheng Gu, Zheng-Xin Liu, Xiao-Gang Wen, Phys. Rev. B \textbf{87}, 155114 (2013). 

\bibitem{Wilczek}F. Wilczek , Phys. Rev. Lett. \textbf{49}, 957 (1982).

\bibitem{Bartolomei}
H. Bartolomei, M. Kumar, R. Bisognin, A. Marguerite, J.-M. Berroir , E. Bocquillon, B. Plaçais, A. Cavanna, Q. Dong, U. Gennser, Y. Jin, and G. Fève, Science \textbf{368}, 173 (2020)

\bibitem{Halperin} B. I. Halperin, Phys. Rev. Lett. \textbf{52}  1583  (1984).
%%%%% quantum computation  
\bibitem{Nayak} Chetan Nayak, Steven H. Simon, Ady Stern, Michael Freedman, and Sankar Das Sarma, Rev. Mod. Phys. \textbf{80}, 1083 (2008).

\bibitem{ayukitaev} A. Yu. Kitaev, 
Annals of Physics, \textbf{303}, 2 (2003).

\bibitem{Semeghini} 
G Semeghini, H Levine, A Keesling, S Ebadi, T T Wang, D Bluvstein, R Verresen, H Pichler, M Kalinowski, R Samajdar, A Omran, S Sachdev, A Vishwanath, M Greiner, V Vuleti\'{c}, M D Lukin, Science \textbf{374}, 1242 (2021).


%%%%%%%%%% Some other evidence of spinons 

\bibitem{wzhu} W. Zhu, S.-S. Gong, and D. N. Sheng, PNAS \textbf{116}, 5437 (2019).

\bibitem{yfu} Y. Fu, M.-L. Lin, L. Wang, Q. Liu, L. Huang, W. Jiang, Z. Hao, C. Liu, H. Zhang, X. Shi, J. Zhang, J. Dai, D. Yu, F. Ye, P. A. Lee, P.-H. Tan and J.-W. Mei, Nat. Commn. \textbf{12}, 3048 (2021).

\bibitem{mpotts} M. Potts, R. Moessner, and O. Benton, Phys. Rev. Lett. \textbf{133}, 226701 (2024).

%%%%%%%%%%%  photon AGF  


\bibitem{Cheng}D. Cheng,  K. Wang, R. -C. Charles  E. Lustig, Olivia Y. Long, Heming Wang and Shanhui Fan {Nature} \textbf{637}, 52 (2025). 

\bibitem{Umucalilar} R. O. Umucalilar and I. Carusotto, Phys. Rev. A \textbf{84}, 043804 (2011).

\bibitem{Kejie} Kejie Fang, Zongfu Yu and Shanhui Fan, Nature Photonics \textbf{6}, 782 (2012).

%%%%%%%%%%%  phonon AGF  

\bibitem{Mathew} John P. Mathew, Javier del Pino, and Ewold Verhagen, Nature Nanotechnology \textbf{15}, 198–202 (2020).

\bibitem{xwen}X. Wen,  C. Qiu, Y.  Qi, Y. Qi, M. Ke, F. Zhang and Z. Liu, \textit{Nat. Phys.} \textbf{15}, 352 (2019).

\bibitem{Pengtao} P. Lai, J. Wu, Z. Pu, Q. Zhou, J. Lu, Hui Liu, W. Deng, H. Cheng, S. Chen, Z. Liu, Phys. Rev. Applied \textbf{21}, 044002 (2024). 

\bibitem{liluo} Li Luo, H.-X. Wang, Z.-K. Lin, B. Jiang, Y. Wu, F. Li and J.-H. Jiang, Nature Materials \textbf{20}, 794 (2021).

\bibitem{qiyun} Q. Ma, Z. Pu, L. Ye, J. Lu, X. Huang, M. Ke, H. He, W. Deng, and Z. Liu, Phys. Rev. Lett. \textbf{132}, 066601 
 (2024).

\bibitem{Zhaoju} Zhaoju Yang, Fei Gao, Xihang Shi, Xiao Lin, Zhen Gao, Yidong Chong, and Baile Zhang, Phys. Rev. Lett. \textbf{114}, 114301 (2015).

%%%%%%%%%%% higher-order topology 

\bibitem{Schindler} F. Schindler, A. M. Cook, M. G. Vergniory, Z. Wang, S. S. P. Parkin, B. A. Bernevig, and T.  Neupert, Sci. Adv.  \textbf{4}, eaat0346 (2018).

\bibitem{Schindler2} F. Schindler, Z. Wang, M. G. Vergniory, A. M. Cook, A. Murani, S. Sengupta, A. Y. Kasumov, R. Deblock, S. Jeon, I.Drozdov, H. Bouchiat, S. Gu\'{e}ron, A. Yazdani, B. Andrei Bernevig and T. Neupert, Nature Physics \textbf{14}, 918–924 (2018).

\bibitem{Christopher} Christopher W. Peterson, Tianhe Li, Wladimir A. Benalcazar, Taylor L. Hughes, and Gaurav Bahl, Science \textbf{368}, 1114(2020).

\bibitem{haoran} Haoran Xue, Yahui Yang, Fei Gao, Yidong Chong and Baile Zhang, Nature Materials \textbf{18}, 108 (2019).

\bibitem{congchen} C. Chen, X.-T. Zeng, Z. Chen, Y. X. Zhao, X.-L. Sheng, and S. A. Yang, Phys. Rev. Lett. \textbf{128}, 026405 (2022).

%%%%%%%%%%%%  Rydberg
\bibitem{Maskara}N. Maskara, S. Ostermann
J. Shee, M. Kalinowski, A. M. Gomez, R. A.
Bravo, D. S. Wang, A. I. Krylov,  N. Y. Yao, M. Head-Gordon, M. D. Lukin,
and S. F. Yelin,  Nat. Phys. \textbf{21}, 289 (2025). 

\bibitem{Bluvstein}
D. Bluvstein, A. Omran, H. Levine, A. Keesling, G. Semeghini, S. Ebadi, T. T. Wang, A. A. Michailidis, N. Maskara, W. W. Ho, S. Choi , M. Serbyn, M. Greiner, V. Vuleti\'{c}, M. D. Lukin, Science \textbf{371}, 1355-1359 (2021).

%%%%%%%%%%%%%%%%%%%%%%%%

\bibitem{Kim} Se Kwon Kim, H\'{e}ctor Ochoa, Ricardo Zarzuela, and Yaroslav Tserkovnyak, Phys. Rev. Lett. \textbf{117}, 227201 (2016).

\bibitem{Habel} Jonas Habel, Alexander Mook, Josef Willsher, and Johannes Knolle, Phys. Rev. B \textbf{109}, 024441 (2024).

\bibitem{sup} See supplemental materials for pertinent technical details on relevant proofs and derivations, which includes Ref. \cite{Habel, Haegeman,Ulrich02,Fishman,Naveen,Roushan,dwwang,Barredo,Barthel,Ren,White,Ulrich01}

\bibitem{Ulrich02} U. Schollw\"{o}ck, Ann. Phys. \textbf{326}, 96 (2011)

\bibitem{Haegeman} J. Haegeman, J. I. Cirac, T. J. Osborne, I. Pi\v{z}orn, H. Verschelde,and F. Verstraete, Phys. Rev. Lett. \textbf{107}, 070601(2011).

\bibitem{Fishman} M. Fishman, S. R. White, E. Miles Stoudenmire, 	SciPost Phys. Codebases 4 (2022). 

\bibitem{Barthel} T. Barthel, U. Schollw\"{o}ck, and S. R. White, Phys. Rev. B \textbf{79}, 245101 (2009).

\bibitem{Ren} J. Ren, J. Sirker,  Phys. Rev. B \textbf{85}, 140410 (2012).

\bibitem{White} S. R. White, Phys. Rev. B \textbf{48}, 10345 (1993).

\bibitem{Ulrich01} U. Schollw\"{o}ck, Rev. Mod. Phys. \textbf{77}, 259 (2005).
%%%%%%%%%%%% gauge field ground state topology 
\bibitem{lbshao} L. B. Shao, Q. Liu, R. Xiao, Shengyuan A. Yang, and Y. X. Zhao, Phys. Rev. Lett. \textbf{127}, 076401 (2021).

\bibitem {yxzhao1} Y. X. Zhao, Cong Chen, Xian-Lei Sheng, and Shengyuan A. Yang, Phys. Rev. Lett. \textbf{126}, 196402 (2021)

\bibitem{haoranx} Haoran Xue, Zihao Wang, Yue-Xin Huang, Zheyu Cheng, Letian Yu, Y. X. Foo, Y. X. Zhao, Shengyuan A. Yang, and Baile Zhang, Phys. Rev. Lett. \textbf{128}, 116802 (2022).

\bibitem{zychen} Z. Y. Chen, Z. Zhang, S. Y Yang, Y. X. Zhao, Nature Communications \textbf{14}, 743 (2023).

\bibitem{Luscher} M. Lüscher, \textit{Commun.Math. Phys.} \textbf{85}, 39 (1982). 

%%%%%%%%%%%%%%%%

\bibitem{dhlee} D. -H.  Lee, G. -M. Zhang, and T. Xiang, Phys. Rev. Lett. \textbf{99}, 196805 (2007).

\bibitem{ykitaev} A. Yu. Kitaev,  Phys. Usp. \textbf{44}, 131 (2001).

\bibitem{Jackiw} R. Jackiw and C. Rebbi, Phys. Rev. D \textbf{13}, 3398 (1976).

\bibitem{Schrieffer} W. P. Su, J. R. Schrieffer, and A. J. Heeger, Phys. Rev. Lett. \textbf{42}, 1698 (1979).

\bibitem{footnoteadd}  The first four excited states are topologically trivial, describing spin excitations along the flux line (Sec. 6 of  Supplemental Materials). 

\bibitem{yingran} Ying Ran, Ashvin Vishwanath, and Dung-Hai Lee, Phys. Rev. Lett. \textbf{101}, 086801 (2008).

\bibitem{note1} The local tuning knob $B_i$ is kept as before on the sites around the flux line I and II.

%%%%%%%%%%%%% General Floquet theory review  
\bibitem{Eckardt} A. Eckardt, Rev. Mod. Phys. \textbf{89}, 011004 (2017).

\bibitem{Bukov} M.  Bukov, L. D'Alessio, and A. Polkovnikov, Advances in Physics 
\textbf{64} (2), 139–226 (2015).
%%%%%%%%%%%%%%%%

\bibitem{Anisimovas} A. Eckardt and E. Anisimovas, New J. Phys. \textbf{17}, 093039 (2015).

\bibitem{wenbook} X. G. Wen, \textit{Quantum field theory of many-body systems}, Oxford University Press (2004).

%%%%%%%%%%%%%% Rydberg XXZ 

\bibitem{Whitlock} S. Whitlock, A. W. Glaetzle, and P.  Hannaford, J. Phys. B: At. Mol. Opt. Phys. \textbf{50} 074001 (2017).

\bibitem{Kunimi} M. Kunimi, T. Tomita, H. Katsura, Y. Kato, Phys. Rev. A \textbf{110}, 043312 (2024).

%%%%%%%%%%%%% Rydberg control

\bibitem{Naveen} N. Nishad, A. Keselman, T. Lahaye, A. Browaeys, and S. Tsesses, Phys. Rev. A \textbf{108}, 053318 (2023).

\bibitem{Roushan} P. Roushan, C. Neill, A. Megrant, Y. Chen, R. Babbush, R. Barends, B. Campbell, Z. Chen, B. Chiaro, A. Dunsworth, \textit{et al}., Nat. Phys. \textbf{13}, 146 (2017).

\bibitem{dwwang} D.-W. Wang, C. Song, W. Feng, H. Cai, D. Xu, H. Deng, H. Li, D. Zheng, X. Zhu, H.Wang, S.-Y. Zhu and M. O. Scully, Nat. Phys.\textbf{15}, 382 (2019).
	
\bibitem{Barredo} S. de L\'{e}s\'{e}leuc, D. Barredo, V. Lienhard, A. Browaeys, and T. Lahaye, Phys. Rev. Lett. \textbf{119}, 053202 (2017).

%\bibitem{note2} Note that the Ising interaction $K^{\prime}$  here merely stabilizes the desired ferromagnet ground state, and does not affect our obtained results.

\bibitem{Peyronel} T. Peyronel, O. Firstenberg, Q. -Y. Liang, S. Hofferberth, A. V. Gorshkov, T. Pohl, M. D. Lukin and V. Vuleti\'{c}, Nature \textbf{488}, 57–60 (2012).

%%%%%%%%%%%%%%%%% Rydberg measurement 
\bibitem{Endres} M. Endres, H. Bernien, A. Keesling, H. Levine, E. R. Anschuetz, A. Krajenbrink,  C. Senko, V. Vuletic, M. Greiner, Mi. D. Lukin, Science \textbf{354}, 1024 (2016).

\bibitem{Okuno} D. Okuno, Y. Nakamura, T. Kusano, Y. Takasu, N. Takei, H. Konishi, Y. Takahashi, J. Phys. Soc. Jpn. \textbf{91}, 084301 (2022)

\bibitem{Kwona} K. Kwon, K. Kim, J. Hur, S. Huh, and J. Choi, Phys. Rev. A \textbf{105}, 033323 (2022).

%%%%%%%%%%%%%%%%%


\bibitem{rwang} Rui Wang, Tigran A. Sedrakyan, Baigeng Wang, Lingjie Du and  Rui-Rui Du, Nature \textbf{619}, 57 (2023).

\bibitem{Serbyn} Maksym Serbyn, Dmitry A. Abanin and Zlatko Papi\'{c}, Nat. Phys. \textbf{17}, 675 (2021).



\end{thebibliography}
\end{document}